\documentclass[twocolumn,showpacs,preprintnumbers,amsmath,amssymb,showkeys]{revtex4}

\usepackage{graphicx}
\usepackage{dcolumn}
\usepackage{bm}
\usepackage{color}

\begin{document}

\preprint{Superlattices and Microstructures, Volume 25, Issues 1-2, January 1999, Pages 213-219}

\title{Study and characterization by magnetophonon resonance of the energy structuring in GaAs/AlAs quantum-wire superlattices}

\author{T. Ferrus \footnote{Present address : Hitachi Cambridge Laboratory, J. J. Thomson Avenue, CB3 0HE, Cambridge, United Kingdom}}
\email{taf25@cam.ac.uk}
\author{B. Goutiers}
\author{L. Ressier}
\author{J. P. Peyrade}
\affiliation {Institut National des Sciences appliqu\'ees, 135 Avenue de Rangueil, 31077 Toulouse, France}
\author{J. Galibert}
\affiliation{Laboratoire National des Champs Magn\'etiques Puls\'es, UMR 5842, 143 Avenue de Rangueil
31400 Toulouse, France}
\author{J. A Porto}
\author{J. S\'anchez-dehesa}
\affiliation{Departamento de F\'isica Te\'orica de la Materia Condensada, Universidad Aut\'onoma de Madrid, Madrid, 28049, Spain}

\date{\today}

\begin{abstract}

We present the characterization of the band structure of GaAs/AlAs quantum-wire 1D superlattices performed by magnetophonon resonance with pulsed magnetic fields up to 35 T. The samples, generated by the \textit{atomic saw method} from original quantum-well 2D superlattices, underwent substantial modifications of their energy bands built up on the X-states of the bulk. We have calculated the band structure by a finite element method and we have studied the various miniband structures built up of the masses $m_t$ and $m_l$ of GaAs and AlAs at the point X. From an experimental point of view, the main result is that in the 2D case we observe only resonances when the magnetic field $\textbf{B}$ is applied along the growth axis whereas in the 1D case we obtain resonances in all magnetic field configurations. The analysis of the maxima (or minima for $\textbf{B}$ // $\textbf{E}$) in the resistivity $\rho_{xy}$ as a function of $\textbf{B}$ allows us to account, qualitatively and semi-quantitatively, for the band structure theoretically expected.

\end{abstract}

\maketitle

Full article available at doi:10.1006/spmi.1998.0640

\end{document}